\documentclass[10pt,twoside,notitlepage,a4paper]{article}
\usepackage{graphicx}
\usepackage{amsmath}
\usepackage{amssymb}

\begin{document}
\title{Avoiding Boundary Effects in Wang-Landau Sampling}
\author{B.J. Schulz, K. Binder, M. M\"uller and D.P. Landau$^{*}$\\ 
{\footnotesize Institut f\"ur Physik, WA331,
Johannes Gutenberg Universit\"at} \\ {\footnotesize Staudinger Weg 7, 
D55099 Mainz, Germany} \\
{\footnotesize $^{*}$Center for Simulational Physics, The
University of Georgia, Athens, Georgia 30602} }
\maketitle
\begin{abstract}
A simple modification of the ``Wang-Landau sampling'' algorithm
removes the systematic error that occurs at the boundary of the
range of energy over which the random walk takes place in the
original algorithm.
\end{abstract}
In two recent papers \cite{WangLandau1,WangLandau2} 
an efficient Monte Carlo procedure was introduced that used a
random walk in energy space to obtain an accurate estimate of the
energy density of states $g(E)$ for classical statistical models.
If this method (now commonly termed ``Wang-Landau Sampling'') is
applied to a restricted energy range, effects at the boundaries of
the energy range come into play, and systematically larger errors
in $g(E)$ at the edges of the sampled energy interval are
observed.  Since the method is of quite general applicability, a
better understanding of these ``edge'' effects could be of
considerable value.  Here, we show how such an enhancement of
errors at the edges can be avoided by a simple modification of the
algorithm.

In Wang-Landau Sampling one accepts trial configurations with
probability $\min(1,g(E)/g(E'))$, where $g(E)$ is the energy
density of states (DOS) and $E$ and $E'$ are the energies of the
current and the proposed configuration, respectively. At each spin
flip trial the DOS is modified $g(E)\rightarrow g(E)\cdot f$ by
means of a modification factor $f$, which is systematically
reduced according to $f\rightarrow f^{1/2}$ whenever the recorded
energy histogram $H(E)$ becomes sufficiently flat that all entries
are within some percentage $\epsilon$ of the average energy
histogram, i.e., $H(E) \geq \epsilon\langle H(E')\rangle_{E'}$ for
all $E$. $H(E)$ is then reset to zero, and the procedure is
repeated until a flat $H(E)$ is achieved using a final
modification factor $f_{final}$. Restricting now the random walk
to some sub-interval of the entire energy range of the system, one
has obviously two basic choices to proceed in case the random walk
is at the border of the considered energy interval and a spin flip
trial would result in an energy outside the specified energy
segment:
\begin{enumerate}
\item Reject the suggested spin flip and do not update $g(E)$ and
the energy histogram $H(E)$ of the current energy level $E$
\label{itm1} \item Reject the suggested spin flip and count the
current energy level once more, i.e., update $g(E)$ and $H(E)$:
$g(E)\rightarrow g(E)\cdot f$ and $H(E)\rightarrow H(E)+1$
 \label{itm2}
\end{enumerate} Method \ref{itm1} was used in Refs. 
\cite{WangLandau1,WangLandau2}
and this led to a systematic underestimation of $g(E)$ at borders
of energy intervals\footnote{In addition, the energy levels
$E_{min}$ and $E_{max}$ of a specified interval
$[E_{min},E_{max}]$ have been updated twice in Refs. 
\cite{WangLandau1,WangLandau2}, 
when visited by the random walk.}. 
This effect was examined for the two-dimensional Ising model, with
linear dimension $L=32$ and for three different ranges of allowed
energies:   $E/(JN) \in [-1.7,-1.2]$, $E/(JN) \in [-1.8,-1.1]$, as
well as $E/(JN) \in [-1.9,-1.0]$.  The results showed that
systematic deviations from the exact DOS occurred only at 
\begin{figure}
\begin{center}
\includegraphics[scale=0.30]{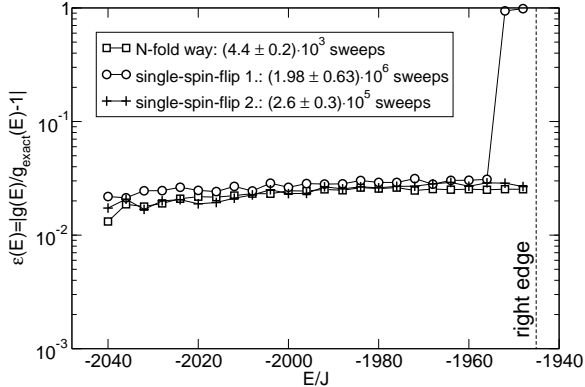}
\caption{Relative error $\varepsilon (E)$ in $g(E)$ for the first
$25$ energy levels of a two-dimensional nearest neighbor Ising
model with linear dimension $L=32$.  Note that the energy scale
was not normalized by the number of spins.  $g(E)$ was obtained by
normalizing with respect to the groundstate, $\varepsilon (E)$ is
an average over $30$ runs. We have used
$f_{final}\simeq\log_{10}(8.09\cdot 10^{-10})$ and
$\epsilon=0.95$.  \label{edges}}
\end{center}
\end{figure}
\begin{figure}
\begin{center}
\includegraphics[scale=0.30]{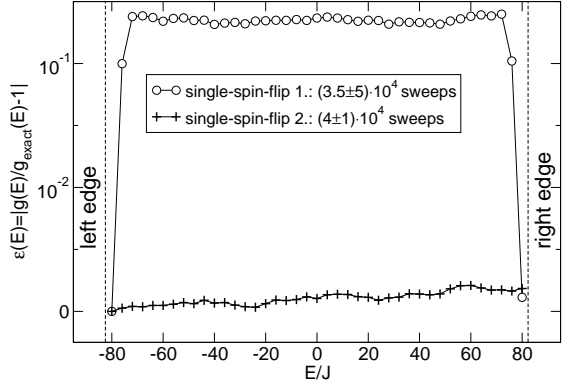}
\caption{Relative error $\varepsilon (E)$ in $g(E)$ for the
interval $E/J \in [-80,80]$ of a two-dimensional nearest neighbor
Ising model with linear dimension $L=32$.  Note that the energy
scale is not normalized by the number of spins.  $g(E)$ was
obtained by normalizing with respect to the exact DOS at the left
edge ($E/J=-80$). $\varepsilon (E)$ is an average over $5$ runs.
We have used $f_{final}\simeq\log_{10}(8.09\cdot 10^{-10})$ and
$\epsilon=0.95$. \label{edges2}}
\end{center}
\end{figure}
the right edges of the energy intervals \cite{WangLandau2}. Since for
the model at hand, this effect only influenced two energy levels
directly at the border, the recipe used was to overlap the
individual intervals over which $g(E)$ was sampled by a sufficient
number of energy levels so that the affected energy levels could
be discarded from each when joining the DOS afterwards. The
asymmetry of this effect can be explained quite simply:  For the
chosen intervals, $g(E)$ has its minimum at the left edge and
increases monotonically as $E$ approaches the right edge. Hence,
during the simulation the random walk is ``pushed'' against the
right edge of the sampled energy range, simply because generating
configurations with energies higher than the right edge energy is
more likely than generating configurations with energies lower
than the boundary energy at left edges. Therefore, for each
interval, a pronounced effect was only visible at the right edge.
In order to demonstrate this we have calculated $g(E)$ for the
first $25$ levels of a $L=32$ two-dimensional Ising model using
single-spin-flip Wang-Landau sampling in both variants (method
\ref{itm1} and \ref{itm2}), as well as N-fold way updates, which
are known not to produce an enhancement of errors at edges. In
\cite{SchulzBinderMueller} the latter algorithm was tested
concerning its behavior at edges against the original
single-spin-flip version, whereby it was misleadingly stated that
boundary effects occur when $g(E)$ at edges is sampled the same
way as inside the energy interval (method \ref{itm2}).  This is
actually incorrect. From the simulation results, depicted in Fig.
\ref{edges}, one clearly sees that method \ref{itm1}, which is
almost identical to the implementation of Wang-Landau
\cite{WangLandau1,WangLandau2}, leads to systematic errors in the
density of states at the right edge (indicated by a dashed line),
where two levels are affected, as described in \cite{WangLandau2}.
When $g(E)$ is sampled according to method \ref{itm2}, no
systematic errors are present. In case the chosen interval is
symmetric around $E=0$, the effect should have the same magnitude
for both edges of the interval, as can be seen from Fig.
\ref{edges2}. Again, no systematical enhancement of errors at
edges is present when the DOS is sampled using method \ref{itm2}.
\section*{Acknowledgments}
This research was partially supported by the Deutsche
Forschungsgemeinschaft (DFG) under grant No. Bi314/16-2, the
Bundesministerium f\"ur Bildung und Forschung (BMBF) under grant
No. 03N6015, and by NSF grant DMR-0094422.
\end{document}